\documentstyle[ApJ,psfig]{article}

\begin{document}

\title{Evidence for an Early High-Energy Afterglow Observed with
BATSE from GRB980923}

\author{T. W. Giblin\altaffilmark{1,2}, J. van Paradijs\altaffilmark{1,3},
C. Kouveliotou\altaffilmark{2,4}, V. Connaughton\altaffilmark{5}, \\
R.A.M.J. Wijers\altaffilmark{6}, M. S. Briggs\altaffilmark{1,2}, 
R. D. Preece\altaffilmark{1,2}, G. J. Fishman\altaffilmark{2}}

\altaffiltext{1}{Department of Physics, University of Alabama in Huntsville,
Huntsville, AL 35899, USA}
\altaffiltext{2}{NASA Marshall Space Flight Center, SD50, Huntsville, AL
35812, USA}
\altaffiltext{3}{Astronomical Institute ``Anton Pannekoek'', University
of Amsterdam, \& Center for High Energy Astrophysics, Kruislaan 403,
1098 SJ Amsterdam, The Netherlands}
\altaffiltext{4}{Universities Space Research Association}
\altaffiltext{5}{NRC, NASA Marshall Space Flight Center, SD50, Huntsville,
AL 35812, USA}
\altaffiltext{6}{Department of Physics and Astronomy, SUNY, Stony Brook,
NY 11794-3800, USA}

\begin{abstract}

In this {\it Letter}, we present the first evidence in the BATSE data for 
a prompt high-energy (25-300 keV) afterglow component from a $\gamma$-ray 
burst (GRB), GRB980923. The event consists of rapid variabilty lasting 
$\sim 40$ s followed by a smooth power law emission tail lasting $\sim 
400$ s. An abrupt change in spectral shape is found when the tail becomes 
noticeable. Our analysis reveals that the spectral evolution in the tail 
of the burst mimics that of a cooling synchrotron spectrum, similar to 
the spectral evolution of the low-energy afterglows for GRBs. This 
evidence for a separate emission component is consistent with the 
internal-external shock scenario in the relativistic fireball picture. 
In particular, it illustrates that the external shocks can be generated 
during the $\gamma$-ray emission phase, as in the case of GRB990123. 

\end{abstract}

\centerline{{\it Accepted: 1999 August 12}}


\section{Introduction}

The distinction between $\gamma$-ray bursts (GRBs) proper and their 
afterglows may be one between emission from shocks internal to the 
relativistic outflow and an external shock with a circumsource medium 
(Rees and M\'{e}sz\'{a}ros 1994; Sari and Piran 1997; M\'{e}sz\'{a}ros 
and Rees 1997). From an observational viewpoint, the relation between 
the burst proper and afterglow emissions is not well understood. However, 
the recent identification of a simultaneous optical counterpart to 
GRB990123 made with ROTSE (Akerloff et al. 1999) has shown that in some 
cases emissions from internal and external shocks can overlap, in agreement 
with the work of Sari and Piran (1999a).

One way to assess this relation has been the comparison of the 1-10 keV
X-ray emission during the burst with that observed in the afterglow. 
Backward extrapolations of the {\it Beppo}SAX X-ray afterglow light curves 
to the decay part of the burst indicate that the late X-ray emission in the 
burst and the afterglow are consistent with a single power law decay curve 
indicating that the tail of the burst evolves continuously into the 
afterglow (Costa et al. 1997; Piro et al. 1998; Nicastro 1998; 
In't Zand 1998). These results appear to be in contradiction with the 
initial disappearance of X-ray emission in GRB780506, which was followed 
$\sim 400$ s later by their reappearance, supporting the case of afterglow 
emission disconnected from the burst (Connors and Heuter 1998). Elsewhere, 
evidence for extended hard $\gamma$-ray emission has been reported for 
$\sim 10\%$ of the bursts detected by PHEBUS (Tkachenko et al. 1995). 
OSSE has provided upper limits on high-energy ($>300$ keV) post-burst 
emission from GRB950421 (Matz et al. 1996). Analysis of a subset of GRBs 
dubbed ``FREDs'' (Fast Rise Exponential Decay) from early in the BATSE 
mission has shown a softening trend in the decay of the burst (Bhat et 
al. 1994). More recently, Burenin et al. (1999) report GRANAT/SIGMA 
observations of a soft $\gamma$-ray tail lasting $\sim 1000$ s.  

We present BATSE observations of GRB980923, which show that after a period 
of $\sim 40$ s of strong variability, the signal abruptly enters a phase of 
smooth decay. During this decay phase, which lasts $\sim 400$ s, the 
properties of the $\gamma$-ray emission are very similar to those of 
afterglows at lower energies. Our results support the idea that, in this 
case, the afterglow emission started during the burst, consistent with the 
theoretical prediction of early high-energy afterglow by Sari and Piran 
(1999b). The spectral evolution of this $\gamma$-ray afterglow indicates 
that it is the early, high-energy part of subsequent X-ray and optical 
afterglow.

\section{GRB980923: BATSE Observations}

On 1998 September 23 at 20:10:47.5 UT BATSE was triggered on the 64 ms 
time scale by a long and intense GRB (trigger 7113), shown in Figure 1. 
The burst was not in the field of view of the {\it Beppo}SAX WFC. 
RXTE/PCA made a single observation in stare mode in response to the 
BATSE Rapid Burst Response (Kippen et al. 1997) system, but observed no 
change in count rate. A full scan of the GRB error circle by RXTE/PCA 
was not possible at the time. 

The event consists of complex variability lasting $\sim$ 40 s, followed 
by a smooth emission tail. The 50-300 keV peak flux of the burst on the 
64 ms time scale is $(1.16 \pm 0.02) \times 10^{-5}$ ergs sec$^{-1}$ 
cm$^{-2}$, ranking twelfth in brightness among all BATSE GRBs. The 
fluence of the burst ($>25$ keV) is $(4.84 \pm 0.02) \times 10^{-4}$ 
ergs cm$^{-2}$, ranking third among all BATSE GRBs. The fluence in the 
tail ($> 40$ s) is $\sim 7\%$ of the total burst fluence. 

\section{Temporal and Spectral Analysis}

Typically GRBs are short enough that prior and post background intervals 
can be chosen so that the behavior of the background during the burst 
emission interval can be estimated through interpolation. In the case of 
GRB980923, however, the duration of the emission tail is unknown, therefore 
routine background subtraction is not applicable. We use as background the 
average of the count rates registered when the spacecraft (CGRO) is 
positioned at the point closest in geomagnetic latitude to that at the time 
of the burst on days before and after the burst trigger, as described in 
Connaughton et al. (1999). Background rates were computed using both CONT 
(16 channel, 2.048 s time resolution) and DISCLA (4 channel, 1.024 s time
resolution) data types for comparison. 

The two detectors which have the highest burst count rates also have 
the smallest source angles to Vela X-1. The Vela pulsations were modeled 
using a 7 term harmonic Fourier expansion centered on a barycentric 
frequency equal to 3.546 mHz, as tabulated from BATSE pulsar monitoring 
(M. Finger, private communication). The model was fit to the background 
subtracted DISCLA data in channel 1 (25-50 keV) during a pre-burst interval 
ranging from $-405$ to $-5$ s and a post-burst interval from 300 to 1000 s, 
having a $\chi^{2}$/dof of 1.12 (1067 dof). Fourier spectra of the residual 
rates in the fitting intervals are consistent with that expected from Poisson 
fluctuations. We also modeled the pulsations in the CONT data (used for
spectral modeling) using the same procedure and time intervals. Fits were 
made in each channel for channel 1 through 10 with an average reduced 
$\chi^{2}$ equal to 0.95. 

\subsection{Temporal Modeling}

We modeled the tail count rates with a power law (PL) of the form 
$A(t-t_{0})^{\beta}$. Separate fits were made in five energy 
ranges (25-50, 50-100, 100-300, 25-300, and $>300$ keV) using the 
DISCLA count rate data from 40 to 298 s after the trigger time. For 
comparison, fits were also made using CONT data, binned in energy to 
match the DISCLA channel boundaries. A series of fits were made in 
each energy range for $t_{0}=[-14.995,32.109]$ at 1.024 s 
intervals using a $\chi^{2}$ minimization algorithm with two free 
parameters, $A$ and $\beta$, while the value of $t_{0}$ was held 
constant for each fit. From the fits in the energy ranges below 300 
keV, we find best-fit (i.e., minimum $\chi^{2}$) values for $\beta$ 
in the range $-1.81$ to $-1.64$ (with typical uncertainties $\pm 0.03$) 
for $t_{0}$ in the range 9.6-12.1 s. For the high-energy ($>300$ keV) 
emission we find $\beta = -1.85 \pm 0.53$ for $t_{0}=10.1$ s 
($\chi^{2}$/dof = 0.2). The $99\%$ confidence intervals in $t_{0}$ 
range from $-4$ to $+18$ s, and do not change much with energy. The 
best fit model in the 25-300 keV range using DISCLA data gives 
$\beta = -1.81 \pm 0.02$ ($\chi^{2}$/dof = 1.22) for $t_{0}=9.6$ s 
(see Figure 2). Both $68.3\%$ (dashed) and $99\%$ (solid) confidence 
intervals for $t_{0}$ are shown in Figure 2 with the 25-300 keV time 
history and the PL models from the range of $t_{0}$ values. The PL fits, 
and the obvious requirement that the PL emission cannot exceed the total 
burst signal, indicate that the PL decay started $\sim 20$-25 s after 
the burst trigger. Note that there is no reason why $t_{0}$ for the 
afterglow should be 0, i.e. the time of trigger, since the (internal 
shock) GRB emission and (external shock) afterglow are physically 
different and thus somewhat independent in timing.

\subsection{Spectral Modeling}

For the spectral modeling we used the Vela-corrected CONT data 
(22-1880 keV) and constructed a single time-integrated spectrum 
for a time interval spanning 40.813 to 102.253 s in the tail. At
the time of observation, the onboard LAD look-up table was in soft 
mode, mapping low energies into higher CONT channels and thus 
providing suitable statistics below 50 keV. To examine the spectral 
evolution in the tail we generated ten spectra with 8.192 s time 
resolution (four 2.048-second CONT bins) from 39.789 to 121.709 s 
in the tail. Although the tail of the burst appears to last beyond 
$\sim$250 s, we selected our source intervals such that the model 
fit parameters are reasonably constrained. The fits were made using 
the forward folding technique. The photon model that best represents 
the data was determined by comparing fits made with a single PL (2 
free parameters) and a smoothly broken power law (SBPL) (4 free 
parameters) using the $\Delta \chi^{2}$ statistic (Band et al. 1997). 
We obtain $\Delta \chi^{2}=67$ with a chance probability of obtaining 
a value equal to or greater than $\Delta \chi^{2}$ of $\sim 10^{-15}$. 
Therefore, we used the SBPL to model the spectra for GRB980923.

The parameters of the SBPL are the low- and high-energy photon indices 
$\alpha_{\rm low}$ and $\alpha_{\rm high}$, and the break energy $E_{\rm b}$. 
From the time-integrated fit, we obtain $\alpha_{\rm low}=-1.66 \pm 0.04$, 
$\alpha_{\rm high}=-2.20 \pm 0.11$, and $E_{\rm b} = 203 \pm 59$ keV 
($\chi^{2}/{\rm dof} = 1.06$). The GRB function (Band et al. 1993) was 
also fit ($\chi^{2}/{\rm dof} = 1.07$) giving nearly identical parameter 
values ($\alpha_{\rm GRB}=-1.59 \pm 0.05$, $\beta_{\rm GRB}=-2.192 \pm 0.14$,
$E_{\rm p}=271 \pm 52$), similar to values commonly found in GRB spectra 
(Preece et al. 1999). However, the GRB function proved to be less robust 
than the SBPL in the time resolved fits. From the ten fits made in the 
tail, we find that $\alpha_{\rm low}$ and $\alpha_{\rm high}$ do not vary 
significantly, with weighted average values $-1.68 \pm 0.03$ and 
$-2.09 \pm 0.07$, respectively. The break energy was held fixed in 
several of these fits to allow meaningful estimates of the photon 
indices. Therefore, we repeated the ten fits keeping $\alpha_{\rm low}$ 
and $\alpha_{\rm high}$ constant at their average values, allowing only 
the break energy and amplitude as free parameters. $E_{\rm b}$ 
decreases with time (see Figure 4), indicating that the spectrum 
retains a constant shape as it shifts in time to lower energies. 
 
Time resolved spectra (2.048 s resolution) during the main burst 
emission were also modeled with a SBPL using CONT data corrected 
for deadtime effects and Vela pulsations. Figure 3 shows the 
evolution of $\alpha_{\rm low}$ and $\alpha_{\rm high}$ during the 
main burst and the tail. An abrupt change in the spectral form is 
quite evident where the main burst emission drops sharply and the 
tail of the burst becomes detectable at $\sim 40$ s. The time 
integrated spectrum of the main burst gives GRB function parameters 
($\alpha_{\rm GRB}=-0.61 \pm 0.01$, $\beta_{\rm GRB}=-2.95 \pm 0.08$,
$E_{\rm p}=364 \pm 3$), typical for GRBs (Preece et al. 1999).

Careful inspection of Figure 3 reveals a softening in the low-energy
index, $\alpha_{\rm low}$, near 14 s, where the burst intensity drops 
to roughly the same level as observed in the tail (see Figure 1 and 2). 
Beyond $\sim 14$ s $\alpha_{\rm low}$ appears to maintain slightly 
steeper values, suggesting an overabundance of low-energy flux that 
was not present (or at least very weak) at the time of the burst 
trigger, which may be attributed to the rise of the afterglow since 
the decay must begin $\sim 20$-25 s after the trigger time.

\section{Discussion}

In the external shock scenario of the relativistic fireball model, 
electrons are accelerated to a PL energy distribution 
$N(\gamma_{\rm e})d\gamma_{\rm e} \propto {\gamma_{\rm e}}^{-p}d\gamma_{\rm e}$
by a forward shock generated when the relativistic shell encounters the 
surrounding medium, where $\gamma_{\rm e}$ is the electron Lorentz factor.
The synchrotron spectrum of a relativistic shock with a PL electron 
distribution can be described by four PL regions segmented according 
to $\nu_{\rm a}<\nu_{\rm c}<\nu_{\rm m}$ in the {\it fast-cooling} regime, 
and $\nu_{\rm a}<\nu_{\rm m}<\nu_{\rm c}$ in the {\it slow-cooling} regime 
(Sari et al. 1998), where $\nu_{\rm a}$ is the self-absorption frequency, 
$\nu_{\rm c}$ is the cooling frequency, and $\nu_{\rm m}$ is the 
characteristic synchrotron frequency. For adiabatic evolution, $\nu_{\rm c}$ 
evolves as $t^{-1/2}$, while $\nu_{\rm m} \propto t^{-3/2}$ (Sari et al. 1996). In the fast-cooling regime, 
the evolution may be radiative, in which case $\nu_{\rm m} \propto t^{-12/7}$. 
Adopting the notation of Sari et al. (1998) and ignoring self-absorption, 
the observed spectral flux in the fast-cooling regime is given by 

\begin{equation}
\label{fast_cooling_spectrum}
F_\nu=\cases{ 
F_{\nu,\rm max} (\nu/\nu_{\rm c})^{-1/2}, & $\nu_{\rm c}<\nu<\nu_{\rm m}$, \cr
F_{\nu,\rm max} \left( \nu_{\rm m}/\nu_{\rm c} \right)^{-1/2} 
\left( \nu/\nu_{\rm m} \right)^{-p/2}, & $\nu>\nu_{\rm m}$. \cr
}
\end{equation}

\noindent Similarly, the flux in the slow-cooling regime can be written
as

\begin{equation}
\label{slow_cooling_spectrum}
F_\nu=\cases{ 
F_{\nu,\rm max} (\nu/\nu_{\rm m})^{-(p-1)/2}, & $\nu_{\rm m}<\nu<\nu_{\rm c}$, \cr
F_{\nu,\rm max} \left( \nu_{\rm c}/\nu_{\rm m} \right)^{-(p-1)/2} 
\left( \nu/\nu_{\rm c} \right)^{-p/2}, & $\nu>\nu_{\rm c}$. \cr
}
\end{equation}

\noindent Three characteristics of the synchrotron spectrum allow us 
to distinguish between the fast- and slow-cooling regimes: 1) the change 
in spectral slope across the break frequency, 2) the time dependence of 
the break frequency, and 3) the relation between the temporal and spectral 
PL indices.

For slow-cooling we expect a change in the spectral slope across $\nu_{\rm c}$ 
equal to $p/2-(p-1)/2=0.5$; for fast-cooling, the change across $\nu_{\rm m}$ 
equals $p/2-1/2$. The value of $p$ can be obtained from 
$p/2=-(\alpha_{\rm high}+1)$, leading to $p=2.4 \pm 0.11$, typical for 
afterglows. The expected change in slope in the fast-cooling regime then 
equals $0.7 \pm 0.11$. From the time-integrated spectral fit, we find a 
slope change of $0.54 \pm 0.12$; from the time-resolved fits we obtain an 
average change in slope of $0.42 \pm 0.09$. These values are in good 
agreement with slow-cooling, but do not exclude fast-cooling very 
strongly.

If we are observing a cooling break then we can expect the break energy 
to evolve as $t^{-1/2}$ (Sari et al. 1996) for adiabatic evolution. The 
time-dependent break energy was modeled with a PL of the form 
$E_{0}(t-t_{0})^{\delta}$ in the same manner as the decay of 
the intensity. We find a PL index $\delta = -0.52 \pm 0.12$ 
($\chi^{2}/{\rm dof} = 1.12$) for $t_{0} = 32.109$ s (see Figure 4);
slightly beyond the $99\%$ confidence interval, but within a $99.9\%$
interval and consistent with the PL decay starting $\gtrsim 20$ s after
the trigger time. To obtain a fit with $\delta=-3/2$ or $\delta=-12/7$, 
corresponding to the two possible behaviors of $\nu_{\rm m}>\nu_{\rm c}$ 
(namely the adiabatic and radiative cases, respectively), would require 
$t_{0}$ to be negative, and is thus excluded by the data. We conclude 
that the break energy corresponds to the cooling frequency, i.e. the 
tail is in the slow-cooling regime.

If we insert the time dependence of $\nu_{\rm c}$ and $\nu_{\rm m}$ into 
the expressions for $F_{\nu}$ and let $\alpha \equiv (p-1)/2$ and 
$\alpha^{\prime} \equiv p/2$, where $\alpha$ and $\alpha^{\prime}$ are
the low- and high-energy PL indices of the energy spectrum, then we have 
$F_{\nu} \propto \nu^{-\alpha} t^{-\case{3}{2}\alpha}$ for 
$\nu<\nu_{\rm c}$ and $F_{\nu} \propto \nu^{-\alpha^{\prime}} 
t^{-\case{3}{2}\alpha^{\prime}+\case{1}{2}}$ for $\nu>\nu_{\rm c}$. 
Therefore, the relationships between the temporal and spectral 
indices are $\beta=-3\alpha/2$ for $\nu<\nu_{\rm c}$ and 
$\beta=-3\alpha^{\prime}/2+1/2=-3\alpha/2-1/4$ for $\nu>\nu_{\rm c}$. 
The values of $\alpha_{\rm low}$ and $\alpha_{\rm high}$ from the time 
integrated spectral fits give an expected values of 
$\beta=-0.99 \pm 0.04$ for the low frequency light curve 
($\nu<\nu_{\rm c}$), and $\beta=-1.3 \pm 0.11$ for the high frequency 
light curve ($\nu>\nu_{\rm c}$). Similar values are derived when the 
average spectral indices are used. Since the PL decay must begin 
later than $20$ s after the burst trigger, we can expect the decay 
index to be shallower than $-1.64$. For $t_{0}=32.109$ s we 
obtain $\beta=-0.98 \pm 0.02$ (25-50 keV), in agreement with the 
value computed from the spectral index and in accordance with the 
value of $t_{0}$ that gives the correct break energy PL index. 
Although this fit does not have the lowest $\chi^{2}$ 
($\chi^{2}/{\rm dof}=1.54$), it does satisfy the requirement that 
the PL can not exceed the total burst count rate. The same scenario 
holds for the high frequency light curve. Our measured value 
$\beta=-1.85 \pm 0.53$ in the high frequency regime ($>300$ keV) is
consistent with $\beta=-1.3$ derived from the spectral index. 

In the fast-cooling regime, two cases must be considered. For 
adiabatic blast wave evolution, we expect 
$F_{\nu} \propto t^{(2-3p)/4} = t^{-1.3}$ above $\nu_{\rm m}$ which
agrees with the measured high-energy decay. However, the expected
behavior below $\nu_{\rm m}$ is $F_{\nu} \propto t^{-1/4}$, which 
strongly disagrees with the measured 25-50 keV PL index by at least 
0.76 for all values of $t_{0}$. For radiative blast waves, the behavior 
above $\nu_{\rm m}$ is $F \propto t^{(2-6p)/7} = t^{-1.8}$, in agreement 
with the data; however, below $\nu_{\rm m}$ we expect $F \propto t^{-4/7}$,
which also strongly disagrees with the data for all values of $t_{0}$. 
In summary, we conclude that fast-cooling scenarios with 
$\nu_{\rm c}<\nu_{\rm m}$ do not agree with the evolution 
of the tail of this burst.

\section{Conclusion}

We find similarities between the properties of GRB980923 and those of
GRB920723 (Burenin et al. 1999). Both exhibit long soft $\gamma$-ray 
tails, however the tail of GRB920723 last longer by a factor 2 and 
decays less steeply ($\beta = -0.7$). Burenin et al. report a change 
in spectral index equal to $\sim 0.8$, which is strikingly similar to 
the abrupt change we observe in $\alpha_{\rm low}$. From our spectral 
analysis we have been able to identify which blast wave evolution 
regime is favored in the production of the high-energy tail.

These results suggest a behavior not unlike that of afterglows observed at 
lower energies. In particular, the evolution of the spectrum in the tail of 
the burst appears to mimic the evolution of a synchrotron cooling break in 
the slow-cooling regime. This implies that the transition from fast to 
slow-cooling (i.e., when $\nu_{\rm m}=\nu_{\rm c}$) can take place on (short) 
time scales comparable to the duration of the burst. If the synchrotron 
emission from external shocks is the source of the ``classical'' afterglow, 
then our analysis provides evidence that, at least in some bursts, the 
afterglow may begin during the $\gamma$-ray emission phase and may reveal 
itself in the GRB light curves.
 
\acknowledgements

We thank Titus Galama and John Horack for comments on this manuscript. J.v.P.
acknowledges support from NASA grant NAG 5-3247.

\begin{figure}
   \begin{minipage}{8.8cm}
      \centerline{\psfig{figure=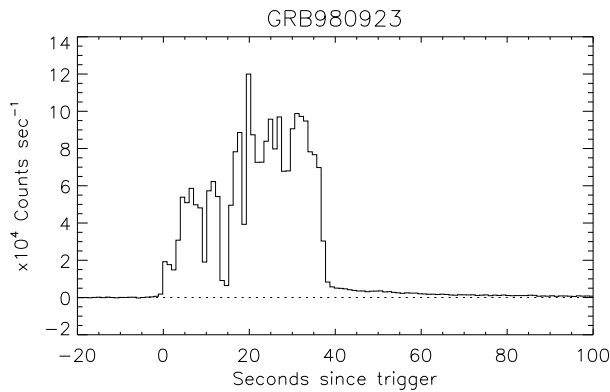,width=\textwidth}}
      \figcaption[fig_1.ps]{DISCLA time history of GRB980923 ($> 25$ 
      keV). \label{figure1}}
   \end{minipage}
\end{figure}

\begin{figure}
   \begin{minipage}{8.8cm}
      \centerline{\psfig{figure=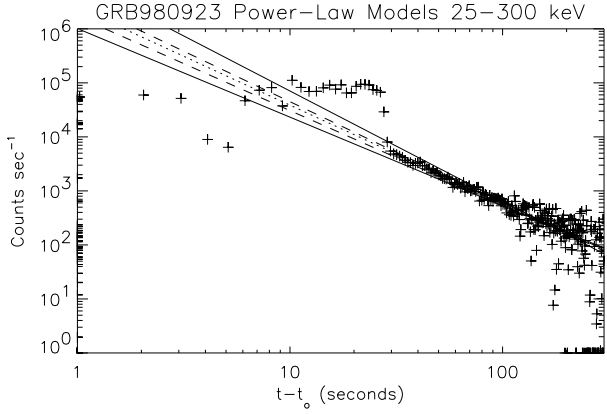,width=\textwidth}}
      \figcaption[fig_2.ps]{DISCLA time history of GRB980923 (25-300 keV) 
      plotted logarithmically with a range of PL models extrapolated 
      backwards in time through the burst proper. The confidence intervals 
      for $t_{0}$ are designated by the dashed (68\%) and solid (99\%) 
      curves. The dotted line is the fit for $t_{0}=9.581$ s 
      ($\beta = -1.81 \pm 0.02$, $\chi^{2}/\rm dof=1.22$). 
      \label{figure2}}
   \end{minipage}
\end{figure}

\begin{figure}
   \begin{minipage}{8.8cm}
      \centerline{\psfig{figure=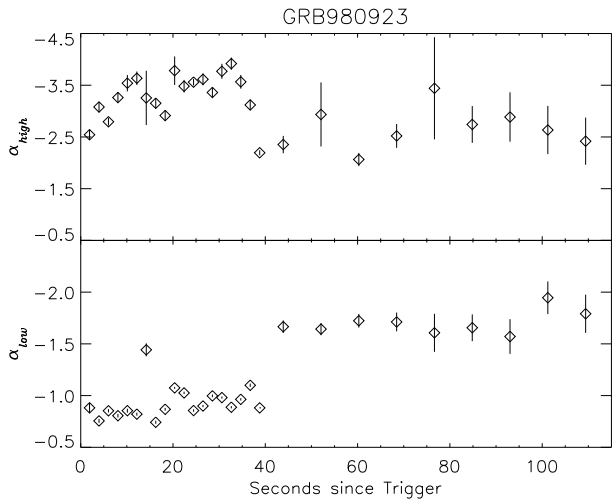,width=\textwidth}}
      \figcaption[fig_3.ps]{SBPL low- and high-energy photon 
      indices (with one-sigma uncertainties) as a function of time 
      since trigger. \label{figure3}}
   \end{minipage}
\end{figure}

\begin{figure}
   \begin{minipage}{8.8cm}
      \centerline{\psfig{figure=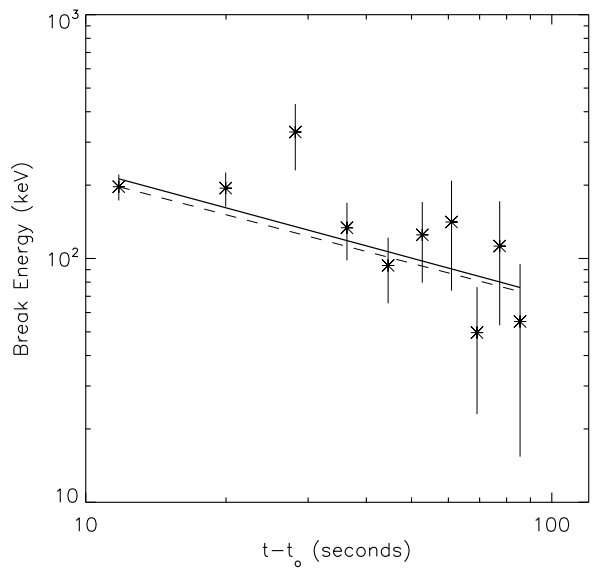,width=\textwidth}}
      \figcaption[fig_4.ps]{Break energy as a function of time since 
      $t_{0}$ on logarithmic axes. The dashed line is the theoretical 
      slope of $-1/2$ for synchrotron cooling. The solid line is the 
      fitted PL slope $-0.52 \pm 0.12$ ($t_{0} = 32.109$, 
      $\chi^{2}/{\rm dof} = 1.12$). \label{figure4}}
   \end{minipage}
\end{figure}

\end{document}